\documentclass[preprint]{aastex}

\usepackage{aas_macros}
\usepackage{amssymb}
\usepackage{amsmath}
\usepackage{natbib}

\begin{abstract}
The co-orbital satellites of Saturn, Janus and Epimetheus, swap radial positions every 4.0 years. Since \textit{Cassini} has been in orbit about Saturn, this has occurred on 21 January in 2006, 2010, and 2014. We describe the effects of this radial migration in the Lindblad resonance locations of Janus within the rings. When the swap occurs such that Janus moves towards Saturn and Epimetheus away, nonlinear interference between now-relocated density waves launches a solitary wave that travels through the rings with a velocity approximately twice that of the local spiral density wave group velocity in the A ring and commensurate with the spiral density wave group velocity in the B ring.
\end{abstract}

\begin{document}
\title{A Traveling Feature in Saturn's Rings}
\author{Morgan E. Rehnberg$^*$, Larry W. Esposito, Zarah L. Brown, Nicole Albers, Miodrag Srem\v cevi\'c, Glen R. Stewart}
\affil{Laboratory for Atmospheric and Space Physics, University of Colorado, 3665 Discovery Drive, Boulder, CO 80303-7820\\
$^*$morgan.rehnberg@colorado.edu}
\maketitle

\section{Introduction}
\label{sec::introduction}
\textit{Cassini} investigations of the rings of Saturn have revealed a rich collection of phenomena both secular and transient.
One of the most important drivers of processes in the rings is the gravitational influence of a subset of Saturn's more than 60 moons. Two of these moons are particularly dynamic: co-orbitals Janus and Epimetheus. Every 4.0 years, they migrate radially and switch their positions relative to the planet. When this migration occurs, so too move the resonance locations between the rings and the two moons. \citet{Tiscareno2006b} have described the effects of this swap on overlapping linear spiral density waves, but stellar occultations observed by \textit{Cassini} at various times during this process have revealed that the nonlinear density waves do not respond in the same way. We investigate that phenomenon here.

\subsection{Resonances and spiral density waves}
Saturn's numerous moons provide for a large number of inner Lindblad resonances within the A and B rings. \citet{Lissauer1982} give the condition for such a resonance as:

\begin{eqnarray}
m(\Omega_p - n) = -\kappa,
\label{eqn::inner_lindblad_resonance}
\end{eqnarray}

where $m\Omega_p = mn_s + k\kappa_s$ is an integer multiple of the angular speed $\Omega_p$ with which the gravitational potential of an orbiting, exterior satellite with no inclination rotates.  It is given in terms of $n_s$, the satellite's mean orbital angular velocity and $\kappa_s$, the radial, or epicyclic, frequency of its orbit. The numbers $n$ and $\kappa$ represent the same quantities for a test particle within the rings.  The coefficients $m$ and $k$ are integers, with $m$ specifying the number of spiral arms the resonance will create.  To first order, $n \approx \kappa,~n_s \approx \kappa_s$ and we can rewrite equation~\ref{eqn::inner_lindblad_resonance} as

\begin{eqnarray}
(m+k)n_s = (m-1)n.
\end{eqnarray}
This leads to the common practice of labeling resonances as $(m+k):(m-1)$.

In order to precisely compute locations of resonances within the rings, however, we must not make approximations like $n_s~\approx~\kappa_s$ and instead compute them separately based on a more complete model of Saturn.  This is necessary because of Saturn's oblateness; if the planet were a perfect sphere, then $n_s~=~\kappa_s$.  We use the method of \citet{Lissauer1982} to compute precise locations for the resonances. Table~\ref{tab::res_params} lists the parameters used in these computations.

\begin{table}
\begin{center}
\begin{tabular} { c c }
Parameter & Value\\
\hline
$GM_\text{Saturn}$ & 37931207.7 km$^3$s$^{-1}$\\
$r_\text{Saturn}$ & 60,330 km\\
$J_2$ & 1.629071$\times 10^{-2}$\\
$J_4$ & -9.3583$\times 10^{-4}$\\
$J_6$ & 8.614$\times 10^{-5}$\\
$J_8$ & -1$\times 10^{-5}$\\
$n_\text{Jan, in}$ & 1.04708751$\times 10^{-4}$ s$^{-1}$\\
$n_\text{Jan, out}$ & 1.04687109$\times 10^{-4}$ s$^{-1}$\\
$n_\text{Epi, in}$ & 1.04736844$\times 10^{-4}$ s$^{-1}$\\
$n_\text{Epi, out}$ & 1.04658870$\times 10^{-4}$ s$^{-1}$\\
\end{tabular}
\caption{The parameters used to compute the resonance locations specified in this work. All except the mean motions of Janus and Epimetheus are taken from \citet{Jacobson2006}.}
\label{tab::res_params}
\end{center}
\end{table}

%\citet{Lissauer1982} give more precise expressions:
%\begin{eqnarray}
%n_s = \frac{GM_p}{r^3}\left[1 + \frac{3}{2}J_2\left(\frac{R_p}{r}\right)^2 - \frac{15}{8}J_4\left(\frac{R_p}{r}\right)^4 + \frac{35}{16}J_6\left(\frac{R_p}{r}\right)^6\right]
%\end{eqnarray}
%and
%\begin{eqnarray}
%\kappa_s = \frac{GM_p}{r^3}\left[1 + \frac{3}{2}J_2\left(\frac{R_p}{r}\right)^2 + \frac{45}{8}J_4\left(\frac{R_p}{r}\right)^4 - \frac{175}{16}J_6\left(\frac{R_p}{r}\right)^6\right],
%\end{eqnarray}
%where $M_p$ is the mass of the planet and $J_i$ are the planet's gravitational coefficients.  Equation~\ref{eqn::inner_lindblad_resonance} can then be solved numerically to compute the precise resonance location.

At many of these resonance locations, features are excited within the rings. \citet{Lin1964} were the first to develop a theory of density waves to explain the spiral structure of many galaxies, thus giving the phenomenon its name. Later, \citet{Goldreich1982} and \citet{Shu1984} were among the first to apply the theory to explain the structure of Saturn's rings.
Linear density waves are raised by weaker resonances as small perturbations on the background surface mass density of the rings. Stronger resonances generate larger perturbations, which can be of the same order as the background density. When this occurs, a nonlinear density wave forms, with sharp peaks (regions of highest density) and shallow troughs (regions of lowest density). The predator-prey model of \citet{Esposito2012} identified strongly-perturbed regions such as nonlinear density waves as a likely site of increased aggregation within the rings.

\subsection{Surface mass density in the A and B rings}
Using images of spiral density waves observed with the \textit{Cassini} Imaging Science Subsystem (ISS), \citet{Tiscareno2007} derive an approximate surface mass density for the A ring of 40 g/cm$^2$.  This surface density appears to increase as one moves farther from the planet.  

\citet{Esposito1983} used a \textit{Voyager} occultation of the Janus 2:1 spiral density wave to estimate a surface mass density for the inner B ring of 70$\pm$10 g/cm$^2$. This is somewhat larger than the results of \citet{Reffet2015}, who use \textit{Cassini} CIRS observations to estimate a value on the order of 40 g/cm$^2$ for the inner B ring, 100 g/cm$^2$ for the middle of the ring and 50 g/cm$^2$ towards the outer boundary. The results of \citet{Hedman2016} bridge this gap by finding that the surface mass density varies within the Janus 2:1 resonance from 69 g/cm$^2$ near resonance to 47 g/cm$^2$ several hundred kilometers exterior. These values are all substantially lower than an estimate that considers the dynamics and aggregation of ring particles by \citet{Robbins2010} of 240-480 g/cm$^2$.

The background surface mass density $\sigma_0$ strongly governs the group velocity $v_g$ of spiral density waves excited at resonance locations, given by $v_g = \frac{\pi G \sigma_0}{\kappa}$, for epicyclic frequency $\kappa$ and gravitational constant $G$ \citep{Toomre1969}. For this study, we assume a surface mass density of 40 g/cm$^2$ for the A ring and as a lower limit for the B ring. We take 70 g/cm$^2$ as an upper limit for the B ring in the vicinity of the Janus 2:1 resonance.

\subsection{Janus and Epimetheus}
Among Saturn's many moons, Janus and Epimetheus are of particular interest because they represent the only pair of co-orbiting satellites in the solar system.  \citet{Dermott1981} showed that this co-orbital configuration leads to each moon traversing a horseshoe orbit about the pair's shared mean orbital radius of 151,450 km in a frame of reference rotating with their mean angular velocity.  This gives a small relative velocity between the two.  Every 4.0 years, the pair approach each other within 15,000 km \citep{Nicholson1992} and exchange orbital angular momentum.  This causes a rapid shift, in which the inner and outer bodies switch position in the course of approximately 100 days.  Each moon is radially shifted in proportion to its relative mass ($m_E/m_J = 0.278$): 20 km for Janus and 80 km for Epimetheus. During \textit{Cassini's} time at Saturn, this has occurred three times: 21 January 2006 (Janus moves inwards), 21 January 2010 (Janus moves outwards), and 21 January 2014 (Janus moves inwards).

Because of this change in the radius of their orbits, special care must be taken when computing the mean motions (and thus the pattern speeds and resonance locations) of these bodies. The semi-major axes given in table~\ref{tab::moon_info} are valid only for times distant from the orbital swap; during the swap, the values evolve continuously. Table~\ref{tab::resonances} lists the computed resonance locations for the first-order Janus and Epimetheus resonances used in this study.

Although the focus of this work is the effect that the Janus/Epimetheus orbital swap has on spiral density waves raised within the rings, \citet{ElMoutamid2016} have also observed that changes in the Janus 7:6 resonance affect the shape of the outer edge of the A ring.

\begin{table}
\begin{center}
\begin{tabular}{ c c c c c c }
Name & $M$ (kg) $^1$& $a$ (km) $^2$& $i$ ($^\circ$) $^3$& $e$ $^3$\\
\hline
Epimetheus &  5.3$\times 10^{17}$ & & 0.351 & 0.0098\\
2002-2006, 2010-2014   &  & 151,410 &  & \\
2006-2010, 2014-2018   &  & 151,490 &  &\\
Janus &  $1.9\times 10^{18}$ & & 0.163 & 0.0068\\
2002-2006, 2010-2014  &  & 151,460 &  & \\
2006-2010, 2014-2018 &  & 151,440 &  &\\
\end{tabular}\\
$^1$\citet{Thomas2010}, $^2$\citet{Jacobson2008}, $^3$\citet{Spitale2006}\\
\caption{Basic properties of Janus and Epimetheus. Values with specified dates refer to 21 Jan of that year and are not accurate in the immediate vicinity (approximately 100 days) of those end points.}
\label{tab::moon_info}
\end{center}
\end{table}

\begin{table}
\center{
\begin{tabular}{ c c c c c c }
Resonance & Time period & $r_{\text{res}}$ (km) & $n_\text{occ}$ & $n_\text{features}$ & $n_\text{empty}$\\
\hline
Epimetheus 6:5 & & & 152 & - & -\\
&2002-2006, 2010-2014 & 134,223 &&&\\
&2006-2010, 2014-2018 & 134,289 &&&\\
Epimetheus 5:4 &&  & 151&-&-\\
&2002-2006, 2010-2014 & 130,660 &&&\\
&2006-2010, 2014-2018 & 130,724 &&&\\
Epimetheus 4:3 && & 142 &-&-\\
&2002-2006, 2010-2014 & 125,228 &&&\\
&2006-2010, 2014-2018 & 125,290 &&&\\
Epimetheus 3:2 && & 132 &-&-\\
&2002-2006, 2010-2014 & 115,922 &&&\\
&2006-2010, 2014-2018 & 115,979 &&&\\
Epimetheus 2:1 && & 103 &-&-\\
&2002-2006, 2010-2014 & 96,216 &&&\\
&2006-2010, 2014-2018 & 96,263 &&&\\
Janus 6:5 && & 152 & 92 & 50\\
&2002-2006, 2010-2014 & 134,265 &&&\\
&2006-2010, 2014-2018 & 134,247 &&&\\
Janus 5:4 &&  & 151 & 110 & 33\\
&2002-2006, 2010-2014 & 130,701 &&&\\
&2006-2010, 2014-2018 & 130,683 &&&\\
Janus 4:3 && & 142 & 127 & 19\\
&2002-2006, 2010-2014 & 125,267 &\\
&2006-2010, 2014-2018 & 125,250 &\\
Janus 3:2 && & 132 & 1 & 104\\
&2002-2006, 2010-2014 & 115,959 &&&\\
&2006-2010, 2014-2018 & 115,943 &&&\\
Janus 2:1 && & 103 & 177 & 5\\
&2002-2006, 2010-2014 & 96,246 &&&\\
&2006-2010, 2014-2018 & 96,233 &&&\\
Mimas 5:3 & & 132,298 & 151 & 7 & 139\\
Prometheus 31:30 & & 136,389 & 146& 0 & 130\\
Prometheus 14:13 & & 132,716 & 153& 0 & 141 \\
\end{tabular}
\caption{Inner Lindblad resonance locations used for this work. The orbit swap of Janus and Epimetheus means that the resonance location shifts on 21 January of the specified years. The number of examined occultations $n_\text{occ}$ does not equal the sum of observed features $n_\text{features}$ and occultations with no observed features $n_\text{empty}$ because some occultations contain multiple features and some have a signal-to-noise ratio too low to resolve the density wave. Because the Epimetheus resonance regions overlap the (much stronger) Janus ones, they were not independently searched for anomalous features.}
\label{tab::resonances}
}
\end{table}

\section{Data}
The data for this investigation consist primarily of stellar occultations observed by the \textit{Cassini} Ultraviolet Imaging Spectrograph (UVIS) High Speed Photometer (HSP). The instrument is described in detail by \citet{Esposito2004}. 

The HSP is a discrete optical train within UVIS designed to make rapid photometric observations of target stars whilst they are being occulted by the phenomenon in question.  With a bandpass of approximately 110-190 nm, HSP is designed to observe bright O and B stars.  A field of view of 6 mrad ensures that the photometer remains evenly illuminated even with non-ideal errors in spacecraft pointing. The photometer can sample at a variable rate from 1 to 8 ms and this rate is fixed for the duration of an observation.  The sampling rate, in addition to occultation geometry and the relative motions of \textit{Cassini} and the rings, gives the spatial resolution of the observation and can be set as rapid as the anticipated data volume permits. For the observations used in this work, an integration period $\Delta t$ of 1 or 2 ms is typical. 

There are three basic components to an HSP observation: the measured stellar photon count during the occultation $I\Delta t$, the observed unocculted stellar photon count $I_0\Delta t$, and the contribution of background sources $b\Delta t$.  For one sampling period, $I$ = $I_0e^{-\tau}+b$, for ring's optical depth $\tau$.  The photon count $I\Delta t$ is the quantity measured as a time series by the photometer.  Ideally, $I_0$ is a constant (ignoring stellar variability), but, in reality, the instrument becomes more sensitive during the course of an observation \citep{Colwell2007}.  This effect is known as ``ramping up.''  Fortunately, this work is not attempting to make absolute photometric measurements and, over short distances (hundreds of kilometers), the effects of ramping up are not large.  Since $b$ should not contain any contribution from the target star's flux, we measure it when the target star is occulted by an opaque region of the rings.

Raw HSP data files contain two vectors: the observed counts $I\Delta t$ and the times at which each data entry was recorded.  The time vector can be converted to vectors of ring radii and ring longitude through knowledge of the position and orientation of the spacecraft, planet, and star. The plane of the rings is defined as the planet's equatorial plane, with longitude measured prograde from the ascending node of Saturn's equatorial plane on Earth's J2000 equator. This information can be obtained from the reconstructed SPICE kernels and geometric calculations are performed with the method described by \citet{Albers2012}. 

We exclude occultations of the stars $\alpha$ Sextantis and $\theta$ Hydrae because the extremely low elevation angle of these observations yields high uncertainty in the geometric solution and a poor signal-to-noise ratio. We estimate the uncertainty of the geometric solution by comparing the location of the inner Encke Gap edge computed by \citet{French1993} with the same feature in each occultation. This yields a 1-$\sigma$ radial uncertainty of 1.17 km, which is smaller than the size of the plotting symbols used in the figures.

By rearranging the above formula for optical depth $\tau$ and correcting for the elevation angle $B$, we find the formula used to convert data counts to normal optical depth:
\begin{eqnarray}
\tau_\perp = \tau\sin B = \ln\left[\frac{I_0}{I-b}\right]\sin B.
\end{eqnarray}
This formula is valid for $I_0 > 0$ (always true) and $I-b > 0$ (not always true).  Because of photon counting statistics, sometimes $b \geq I$ for a given data point.  The natural logarithm is undefined for negative values and unbounded as its argument tends to infinity.  We account for this by placing a floor on $I-b$, namely max($(I-b)\Delta t$,1).  This has the effect of creating a maximum optical depth for observations with low photon counts. Although the presence of self-gravity wakes in the A and B rings may alter $\tau_\perp$ \citep{Colwell2006,Colwell2007,Hedman2007c,Ferrari2009}, the search method described in section~\ref{sec::methods} relies only on relative optical depths $\Delta\tau_\perp$.

\section{Methods}
\label{sec::methods}
\subsection{Search regions}
Each occultation was examined manually over a region spanning 10 km interior to a calculated resonance to 200 km exterior. In the A ring, larger search regions are inhibited by the presence of other resonance locations. In the B ring, the Janus 2:1 density wave becomes very closely spaced. Janus resonances were selected as the primary region of search because they are substantially stronger than the corresponding Epimetheus resonance. In addition, the same region around the Mimas 5:3 (a very strong second-order resonance), the Prometheus 31:30, and the Prometheus 14:13 resonances were searched as a control. The relative strengths of these resonances are given in table~\ref{tab::torque_densities}; all are substantially stronger than those examined by \citet{Tiscareno2006b}.

\begin{table}
\begin{center}
\begin{tabular} { c c }
Resonance & $|T^L_{l,m}|$ (cm$^4$/s$^2$)\\
\hline
Janus 6:5 & 7.90$\times 10^{18}$\\
Prometheus 31:30 & 6.62$\times 10^{18}$\\
Janus 5:4 & 5.09$\times 10^{18}$\\
Janus 4:3 & 2.90$\times 10^{18}$\\
Janus 3:2 & 1.32$\times 10^{18}$\\
Prometheus 14:13 & 1.24$\times 10^{18}$\\
Mimas 5:3 & 6.12$\times 10^{17}$\\
Janus 2:1 & 3.54$\times 10^{17}$\\
\hline
\hline
Janus 11:9 & 4.56$\times 10^{16}$\\
Janus 9:7 	& 1.78$\times 10^{16}$\\
\end{tabular}
\caption{The torque densities computed by \citet{Lissauer1982} for resonances examined in this study (above double line) in comparison to two sample resonances examined by \citet{Tiscareno2006b} (below double line).}
\label{tab::torque_densities}
\end{center}
\end{table}

Some occultations are performed such that the star's track on the rings crosses a given radial location twice. While the radial location may be the same, with few exceptions the azimuthal location in a co-rotating frame is (often substantially) different. Thus we treat the ingress and egress observations as two separate, independent occultations.

\subsection{Selection criteria}
\label{sec::selection_criteria}
The computed optical depth data for each observation were binned to a 500-m resolution to improve the signal-to-noise ratio. Each spiral density wave was then visually searched for anomalous features. A feature was deemed anomalous if it met either of two criteria: (type 1) an optical depth greater than that of the crests of the wave or (type 2) a radial position such that the regularity of the wave was disrupted. To eliminate outliers due to photon counting statistics, a detected feature needed to be at least two data elements wide (approximately 1 km in radial width) and (in the case of type-2 detections) have an optical depth at least half that of the adjacent crests of the wave. A feature is said to have disrupted the regularity of the wave if it broke the monotonic decrease in radial distance between subsequent wave crests when included.

\subsection{Feature characterization}
When an anomalous feature is identified, its position (radius and longitude), peak optical depth, width, and time of observation are recorded. With the exception of the width (measured as FWHM), these quantities are measured from the peak of the feature. In the case of features that break the expected periodicity of the wave (type 2), sometimes, especially later in the wave train, it is difficult to distinguish which of two peaks is anomalous. In these instances, the mean radial location of the two is recorded.

\section{Results}
\label{sec::results}
\begin{figure}
\begin{center}
\includegraphics[width=0.9\textwidth]{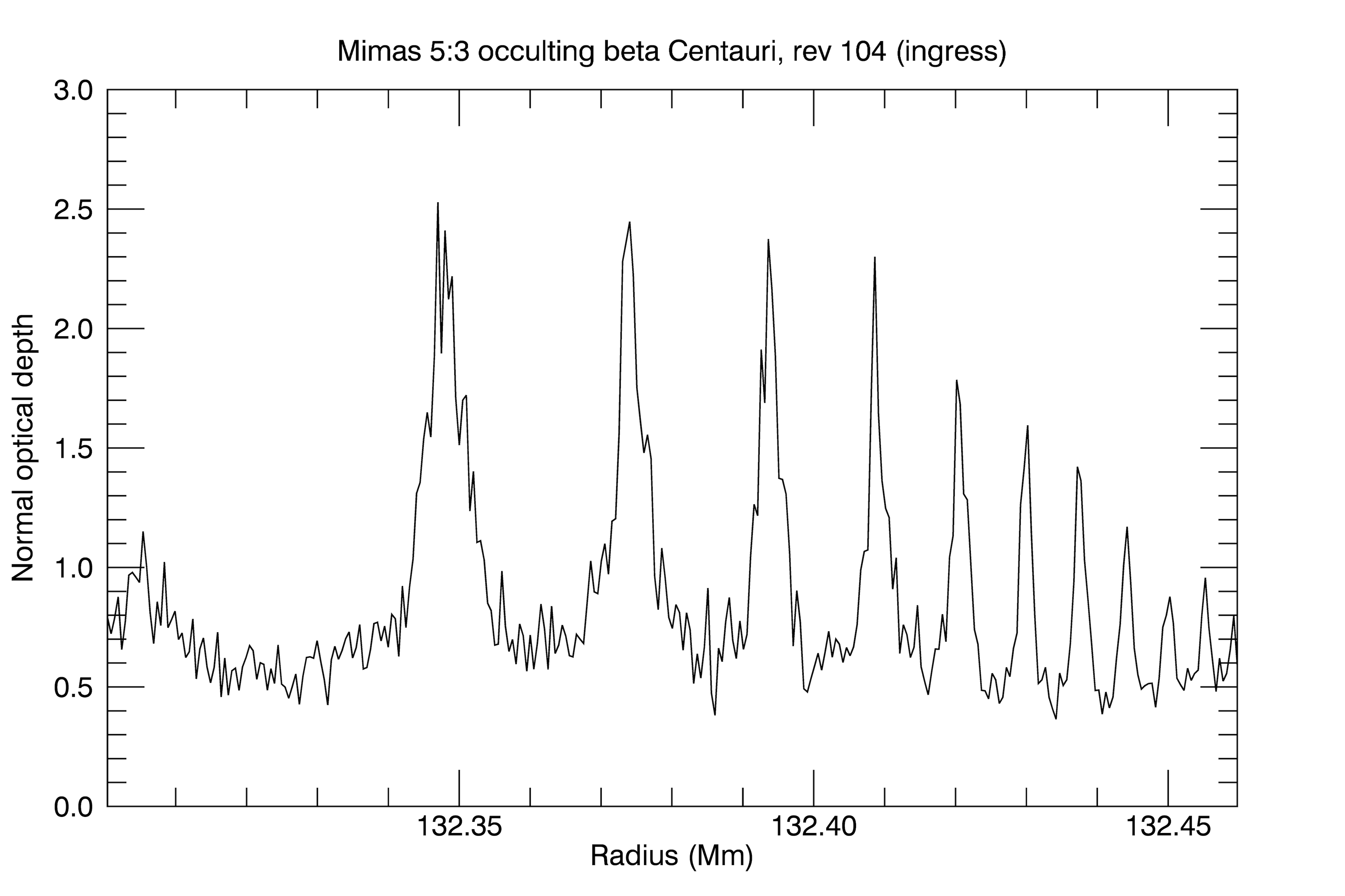}
\caption{A typical nonlinear spiral density wave as seen in an HSP stellar occultation. The data are binned to a radial resolution of approximately 500 meters.}
\label{fig::m53_betcen_104_in}
\end{center}
\end{figure}

\begin{figure}
\begin{center}
\includegraphics[width=0.9\textwidth]{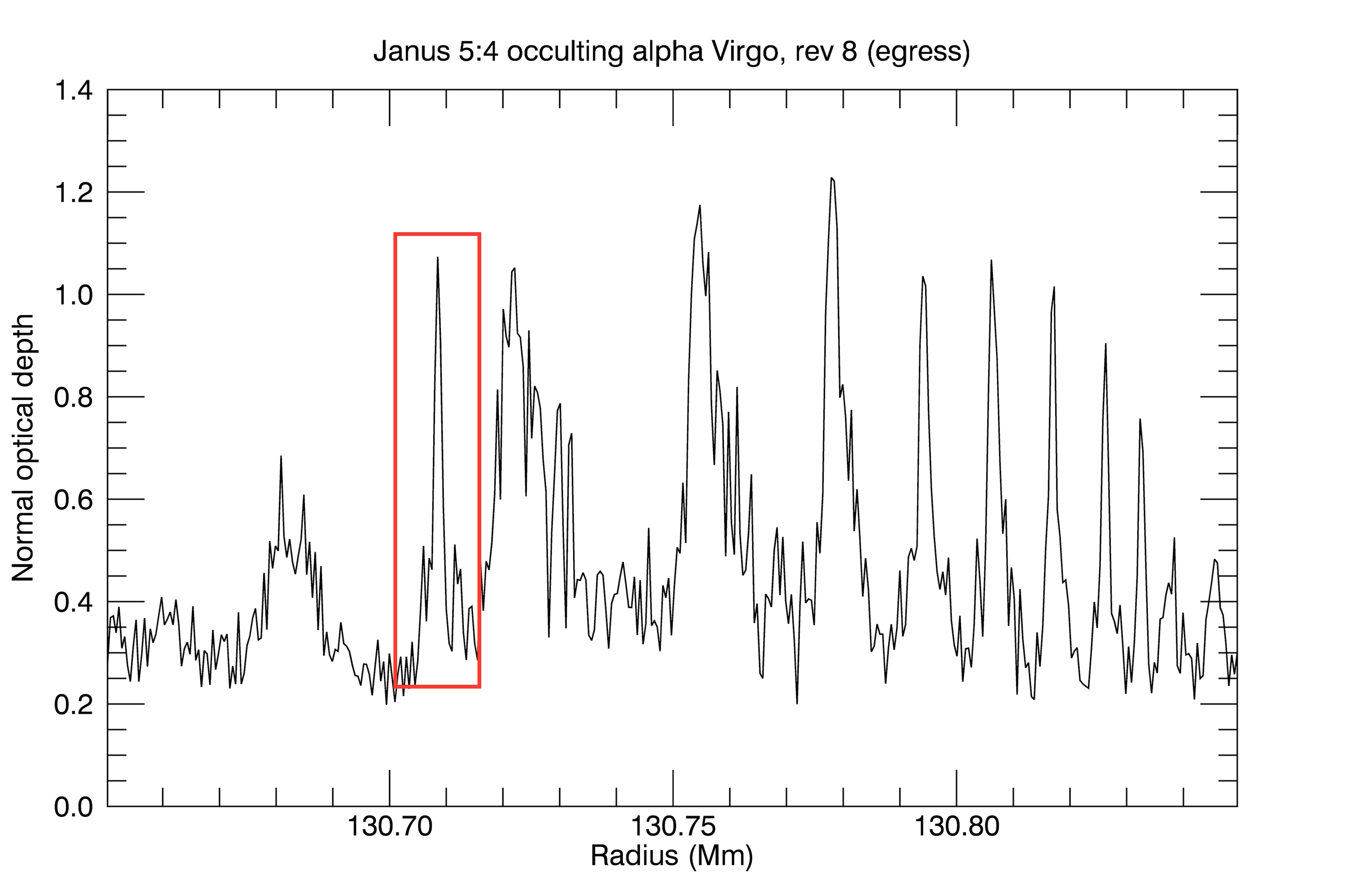}
\caption{An anomalous feature in an HSP stellar occultation of the Janus 5:4 spiral density wave, binned to a radial resolution of approximately 500 meters. The type-2 feature, identified by the red box, clearly breaks the expected regularity of the density wave. The small peak atop the leftmost crest is too narrow to be selected (It would have been a type-1 feature, see section~\ref{sec::selection_criteria}). }
\label{fig::j54_alpvir_8_out}
\end{center}
\end{figure}

The control regions (Mimas 5:3, Prometheus 31:30, Prometheus 14:13) showed few anomalous features matching the criteria outlined in section~\ref{sec::selection_criteria} and no correlation between those was observed. The Janus 3:2 region was too optically thick to distinguish the spiral density wave and thus the selection criteria couldn't be applied. Every other searched region (Janus 6:5, Janus 5:4, Janus 4:3, Janus 2:1) resulted in a substantial number of anomalous features. Figure~\ref{fig::m53_betcen_104_in} illustrates a typical observation without an anomalous feature, while figure~\ref{fig::j54_alpvir_8_out} depicts an occultation containing such a feature.

\begin{figure}
\begin{center}
\includegraphics[width=0.9\textwidth]{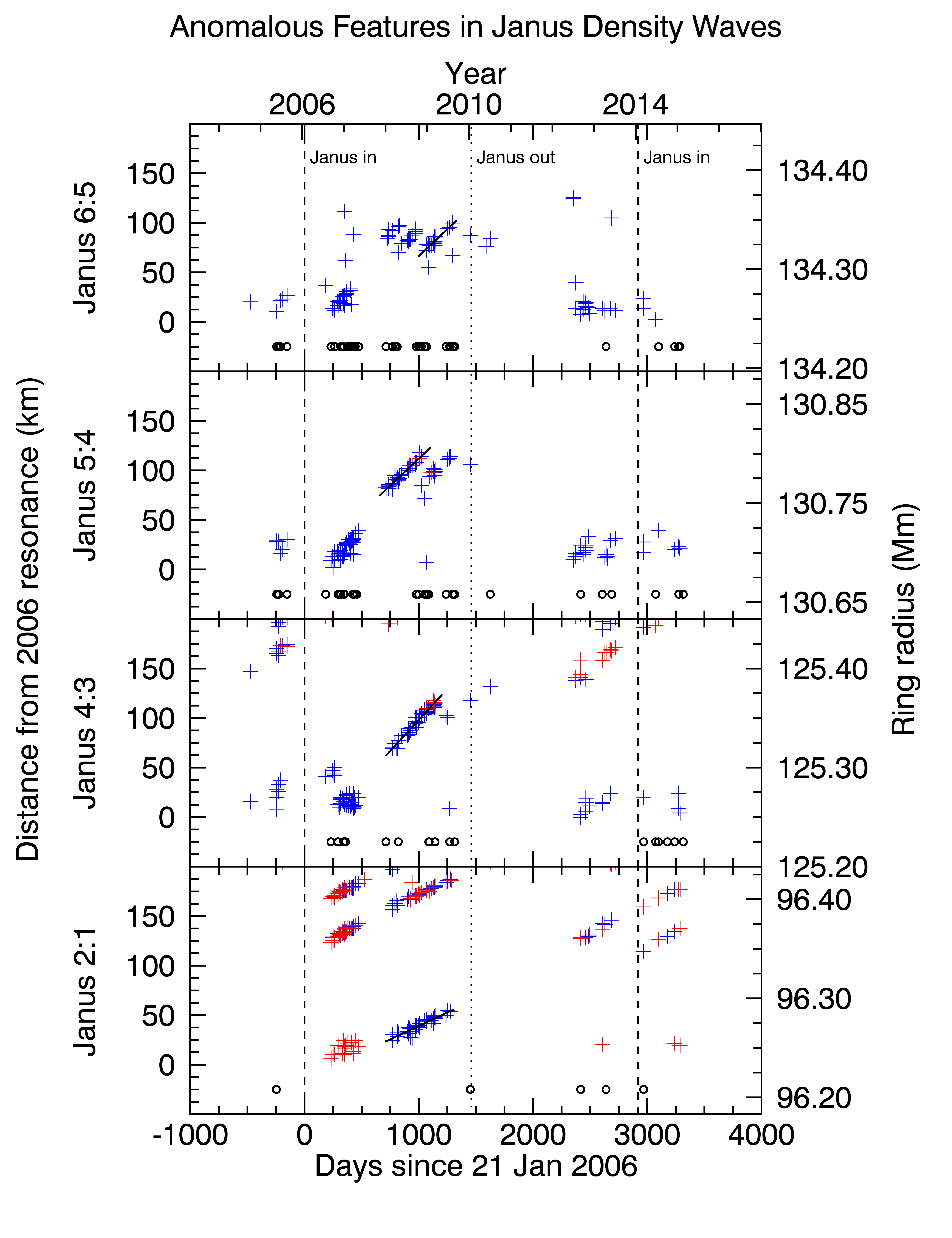}
\caption{Anomalous feature detections in the four Janus resonance search regions. Each detection is indicated by a plus, with blue denoting the radial location of a single peak and red denoting the mean of two neighboring peaks (see section~\ref{sec::methods}). The solid line identifies the points fit to derive the velocities given in table~\ref{tab::velocities}. The circles denote occultations in which no anomalous feature was detected. The 1-$\sigma$ radial uncertainty of 1.17 km is smaller than the plotting symbols. Note the extended gap in observations between mid 2010 and mid 2012 due to the position of \textit{Cassini} in Saturn's ring plane.}
\label{fig::quad_plot}
\end{center}
\end{figure}

For each resonance location, we plot the time a feature was observed against its radial position within the rings. Figure~\ref{fig::quad_plot} depicts the results. A clear trend is visible in each region: as time progresses, the radial locations in which anomalous features are detected move outwards. By fitting a line to a coherent collection of detections in each region (indicated in each figure by a solid line), we can compute the velocity with which a single phenomenon would need to move in order to account for the observed detections. Table~\ref{tab::velocities} lists these fitted values as well as computed group velocities for the local spiral density wave.

Two trends are readily apparent. First, in the A ring, the ``feature velocity'' is approximately twice that of the group velocity in the local spiral density wave, while in the B ring these velocities are roughly commensurate. Second, the number of occultations in which no anomalous features are detected (indicated by the circles near the bottom of each panel in figure~\ref{fig::quad_plot} and tabulated in table~\ref{tab::resonances}) increases as we move to resonances which are farther from Saturn and also stronger. 

\begin{table}
\begin{center}
\begin{tabular}{ c c c}
Resonance & Feature velocity (km/yr) & DW group velocity (km/yr)\\
\hline
Janus 6:5 & 39.8$\pm$1.6 & 21$^1$ \\
Janus 5:4 & 39.6$\pm$0.85 & 20$^1$\\
Janus 4:3 & 45.8$\pm$0.60  & 19$^1$\\
Janus 2:1 &  19.8$\pm$0.55 & 13$^1$,  23$^2$\\
\end{tabular}\\
$^1$assumes $\sigma_0 = 40$ g/cm$^2$, $^2$ assumes $\sigma_0= 70$ g/cm$^2$
\caption{Velocities derived by assuming each anomalous feature is a detection of the same phenomenon and fitting a line as indicated in figure~\ref{fig::quad_plot}. Uncertainty is 1 $\sigma$ with equal weighting of points. Group velocities are computed for the Janus-in configuration.}
\label{tab::velocities}
\end{center}
\end{table}

\begin{figure}
\begin{center}
\includegraphics[width=0.9\textwidth]{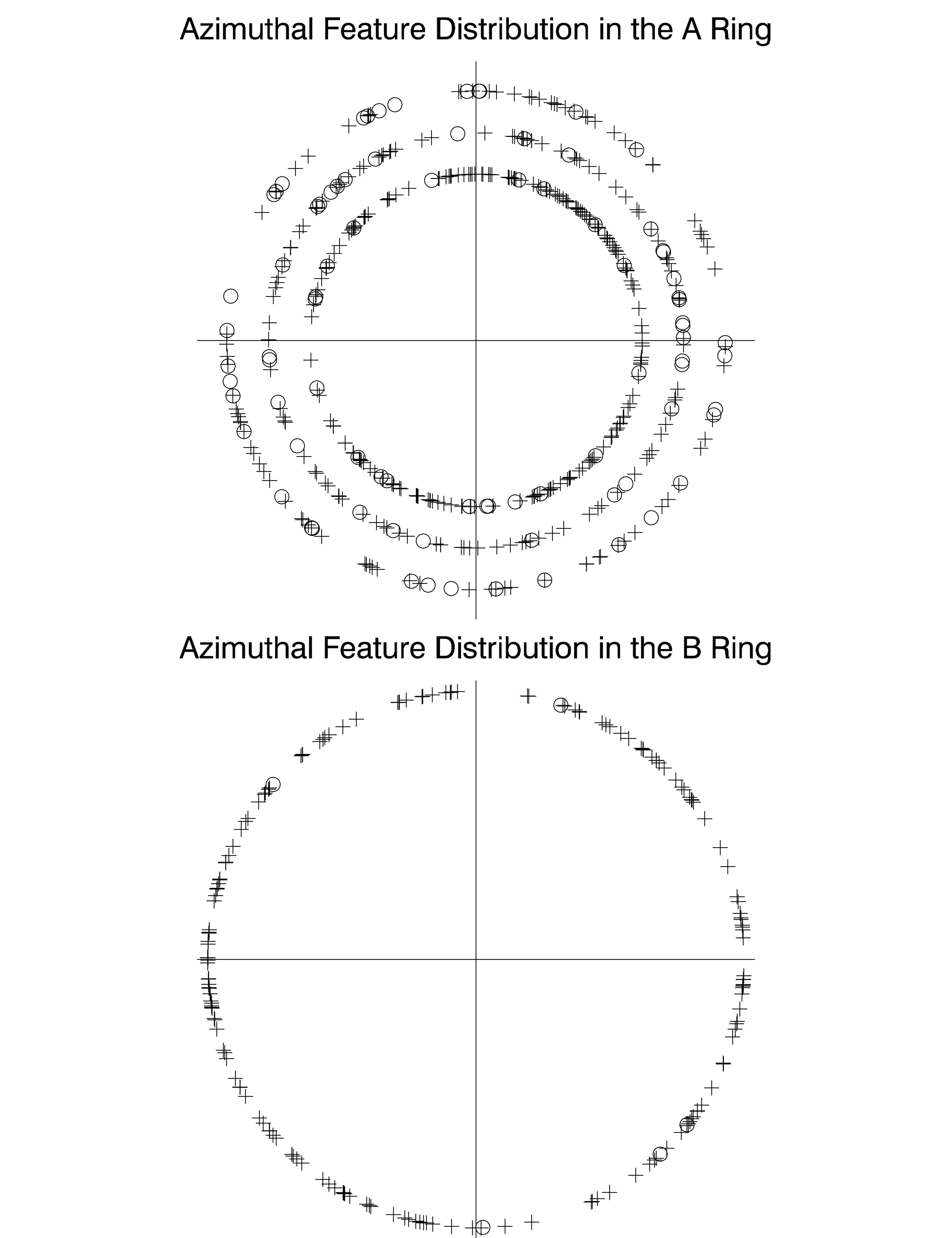}
\caption{The distribution of observed longitudes in a frame co-rotating with the mean motion of the moon. Plusses denote locations in which an anomalous feature was detected. Circles denote locations where no anomalous features were observed. For occultations in which no feature was detected, the longitude at which the occultation crosses the relevant resonance location is used. The A-ring plot is not to radial scale and contains (interior to exterior) the Janus 4:3, Janus 5:4, and Janus 6:5 resonances. The B-ring plot contains the Janus 2:1 resonance..}
\label{fig::longitude}
\end{center}
\end{figure}

No correlation was observed between the detection of an anomalous feature and that occultation's longitude relative to Janus. Figure~\ref{fig::longitude} depicts the longitudinal coverage of our occultations.

\section{Discussion}
We consider two hypotheses with which to explain the observed results: a population of discrete objects and one or more continuous structures. 

\subsection{Hypotheses}
Discrete objects are known to exist in both the main \citep{Tiscareno2006a,Tiscareno2008} and F \citep{Esposito2008,Meinke2012} rings, however none have been observed within the given resonance locations. For such objects to elude detection in ISS images suggests that the size of the object plus any gravitationally-created structure must be smaller than about 500 meters, the approximate best resolution of Narrow-angle Camera (NAC) images. In the Janus 4:3 region, an anomalous feature is detected in 87.6\% of occultations of sufficient quality. We can estimate the quantity of objects present in this location with the formula $N = \frac{2\pi af}{k\bar{d}}$. Setting $a=125,250$ km at the Janus 4:3 resonance location, the observed detection frequency $f$ to 0.876, the azimuthal elongation factor $k$ to 8.5 (derived from \citet{Sremcevic2007}) and the mean object size $\bar{d}$ to 0.5 km, we compute $N~\sim10^5$ objects.

This value seems implausibly large when one considers that these objects appear to migrate out of the resonance region on a four-year timescale. Thus, to sustain such a population would require the generation of  $10^5$ objects every four years. Observed anomalous features are also greater than 500 meters in radial diameter, suggesting they should be detected in ISS imagery. We will return to this subject shortly.

The isotropic nature of anomalous feature detections suggests an alternative hypothesis: that one or more azimuthally-continuous structures are propagating through the rings. Any given occultation will slice the structure at one point along its azimuth at one point in time. As the structure is propagating, occultations at different times will result in detections at different radial locations.

The results in table~\ref{tab::velocities} indicate that the structure propagates about twice as rapidly as the local spiral density wave in the A ring and at approximately the same velocity as the density wave in the B ring. The heretofore-described qualities are consistent with a solitary wave, or soliton, which maintains its inherent structure while passing through background disturbances and has a velocity of propagation set by the initial magnitude of the perturbation. \citet{Norman1978} and \citet{Vukcevic2014} have demonstrated the feasibility of launching a solitary wave within the spiral structure of some galaxies.

An azimuthally-continuous structure should be visible in all occultations that cross its radius, yet we noted in section~\ref{sec::results} that regions that are farther from Saturn contain more observations in which an anomalous feature was not detected (see table~\ref{tab::resonances} for a tabulation of this). This can be accounted for by the increasing number of resonances. Especially in the vicinity of the Janus 6:5 density wave, high-\textit{m} resonances from the small, inner moons (i.e., Prometheus 18:17) become closely spaced. Each launches its own (weak) density wave or, at minimum, perturbs the local surface mass density. This results in increased ``choppiness" in the background optical depth of the ring, making it more difficult to discern an anomalous feature.

\subsection{Janus 4:3 region}
\label{sec::janus_4-3}
\begin{figure}
\begin{center}
\includegraphics[width=0.9\textwidth]{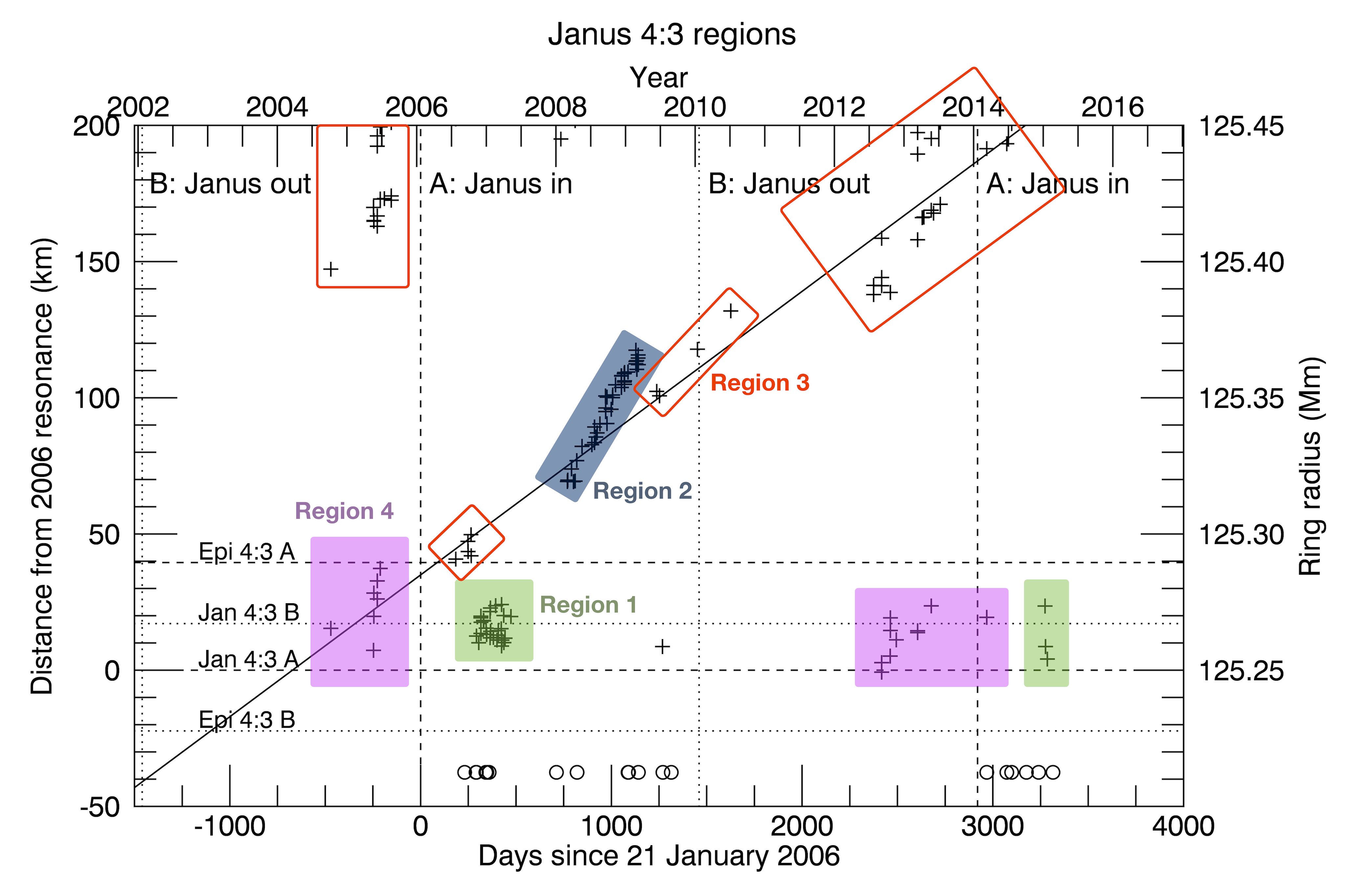}
\caption{An annotated version of a panel from figure~\ref{fig::quad_plot} illustrating the distinct clusters of detections observed. Regions 1 and 4 repeat to the right of the 2014 Janus-in swap. The diagonal line represents the slope with which features moving a the group velocity of the Janus 4:3 spiral density wave would propagate on the plot. See section~\ref{sec::janus_4-3} for a description of how these regions form.}
\label{fig::j43-annotated}
\end{center}
\end{figure}

We observe that not all clusters of anomalous feature detections lie in the included figures such that they imply a propagation velocity consistent with the results of table~\ref{tab::velocities} (region 2 in figure~\ref{fig::j43-annotated}, fitted in figure~\ref{fig::quad_plot}). We take the Janus 4:3 region as a case study. In particular are three additional morphologies of clusters distinct from those described above. Figure~\ref{fig::j43-annotated} depicts these regions, which are grouped by location on the plot, not distinct occultation morphology. The first is a ``blob'' of detections prior to the generation of the solitary wave (region 1). The second is a blob of detections prior to the Janus-in swaps of 2006 and 2014 (region 4). Third is a ``halo'' of detections around the main body of the solitary wave (region 3). We assert the generative mechanism for all these phenomena is the same: constructive interference between density waves. The time, place, and strength of these interactions determines into which phenomenon they manifest. 

\subsubsection{Forming regions 1 and 2}
When the moons swap, the resonances caused by their gravitational perturbations shift correspondingly. The ring, however, cannot respond instantaneously to this new forcing. Instead, the magnitude of the surface mass density perturbation grows with time. When the location of the new resonance falls within the wave train of an existing density wave (generated before the swap) and remains there for some time, strong nonlinear interference can take place. 

As this interference builds, it generates anomalous-looking wave features that produce the blob of detections seen in region 1. When the strength of the perturbation increases past a threshold, it releases the excess negative angular momentum by generating a solitary wave (region 2). 

This explains why the solitary wave is generated during the Janus-in configuration. During this time, the Janus-out-phase Epimetheus wave lies significantly interior to the now-forming Janus-in-phase Janus wave and thus takes more than a year to dissipate from the new Janus resonance location (see figure~\ref{fig::j43-annotated} and table~\ref{tab::resonances} for the resonance locations). For Janus 4:3 this time scale is 1.2 years for the old Epimetheus perturbation traveling at the linear group velocity to reach the nominal new Janus resonance location and longer to pass the most perturbed region. This allows for the maximum perturbation.

\subsubsection{Forming regions 3 and 4}
During the Janus-out configuration, the Janus-in-phase Janus wave, although substantially stronger than the Janus-out-phase Epimetheus wave mentioned above, propagates out of the Janus-out-phase Janus resonance location in less than a year (0.89 yr for Janus 4:3). Thus, no solitary wave forms. Eventually, however, the newly-formed, Janus-out-phase Epimetheus wave reaches the Janus-out-phase Janus resonance and the blob of detections around day 2500 (region 4) can form from weaker nonlinear interference. For the Janus 4:3 wave, with group velocity $v_g \approx 19$ km/yr, this takes 2.1 years. The first region-4 feature after the 2010 swap occurs 2.6 years later. By 2.7 years after the 2002 Janus-out swap, a region-4 feature was also detected, but this represents the earliest available data. Although the swap occurs on 21 Jan, the moons don't immediately reach their final locations. 

The final cluster, the ``halo'' of detections  (region 3) surrounding the main solitary wave path, has actually already been described. It is the region-4 interference propagating for a number of years at a velocity comparable to the group velocity of the local spiral density wave.

\subsubsection{Summary}
In the aftermath of a Janus/Epimetheus orbital swap, the nonlinear interference between the old and new Janus and Epimetheus spiral density waves occurs at different locations and different times. Since this phenomenon is dependent solely on the arrangement of Janus and Epimetheus and the time of observation, we predict that the pattern will repeat itself following the 2014 Janus-in swap. As illustrated in figure~\ref{fig::quad_plot}, we have already observed repetitions of regions 1 and 4. When \textit{Cassini} returns to an inclined orbit in 2016, further occultations should detect a new solitary wave (region 2) and new region-3 propagating interference.

\subsection{Janus 6:5 and Janus 5:4 regions}
The arguments outlined above are also consistent with the Janus 6:5 and 5:4 observations. These locations, though, do differ from Janus 4:3 in region~2. The Janus 5:4 region shows at least two parallel tracks of anomalous detentions in this area, while the Janus 6:5 region shows at least three. We consider two hypotheses for the nature of this difference.

\begin{figure}
\begin{center}
\includegraphics[width=0.9\textwidth]{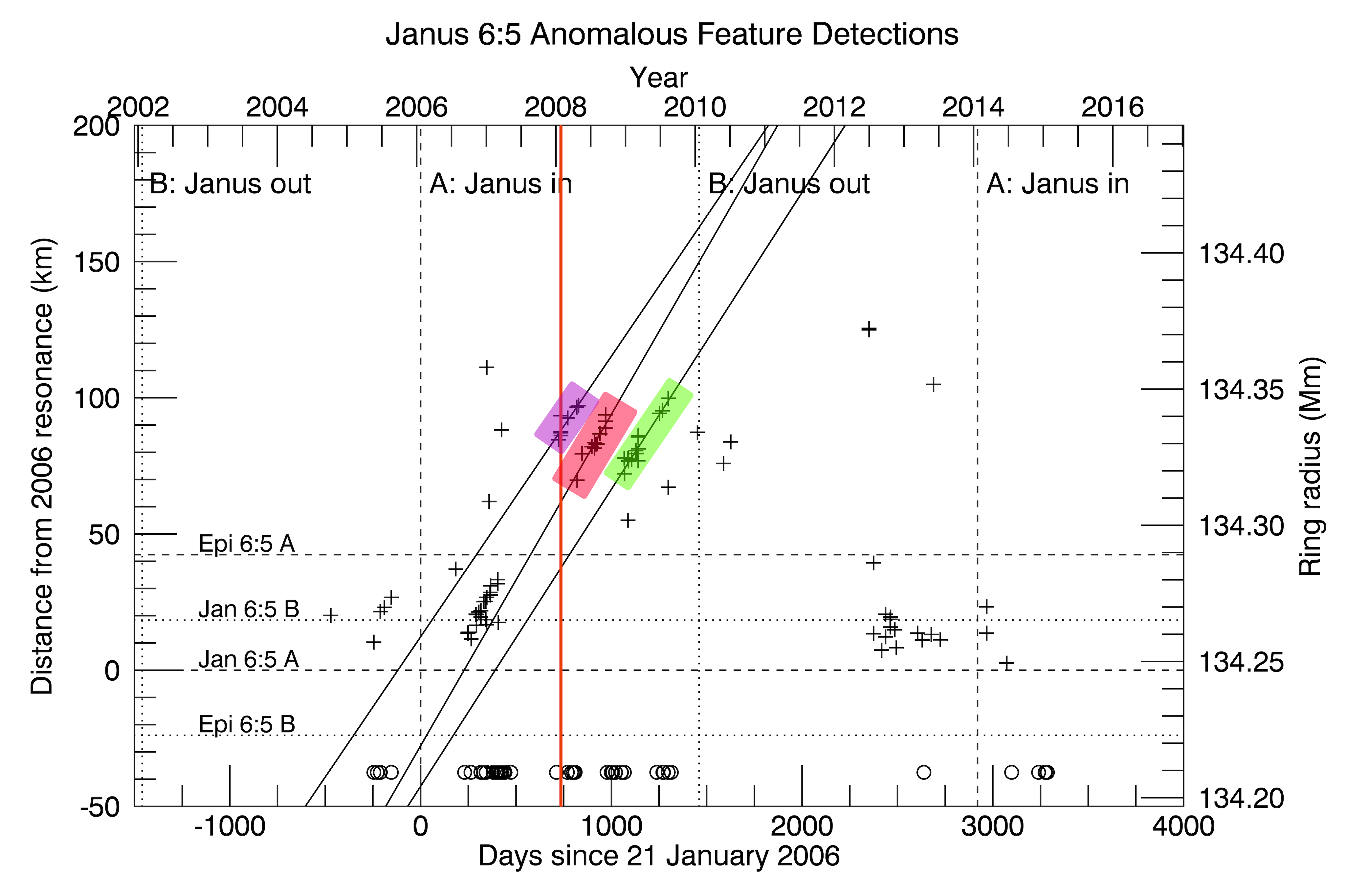}
\caption{The diagonal lines represent fits to the collections of detections indicated by the colored rectangles. The solid, vertical red line indicates the earliest possible date for the Janus-in Janus wave to reach the Janus-in Epimetheus wave. Colors do not correspond to those of figure~\ref{fig::j43-annotated}. The slopes of the lines are tabulated in table~\ref{tab::line_fits}.}
\label{fig::j65-lines}
\end{center}
\end{figure}

\begin{figure}
\begin{center}
\includegraphics[width=0.9\textwidth]{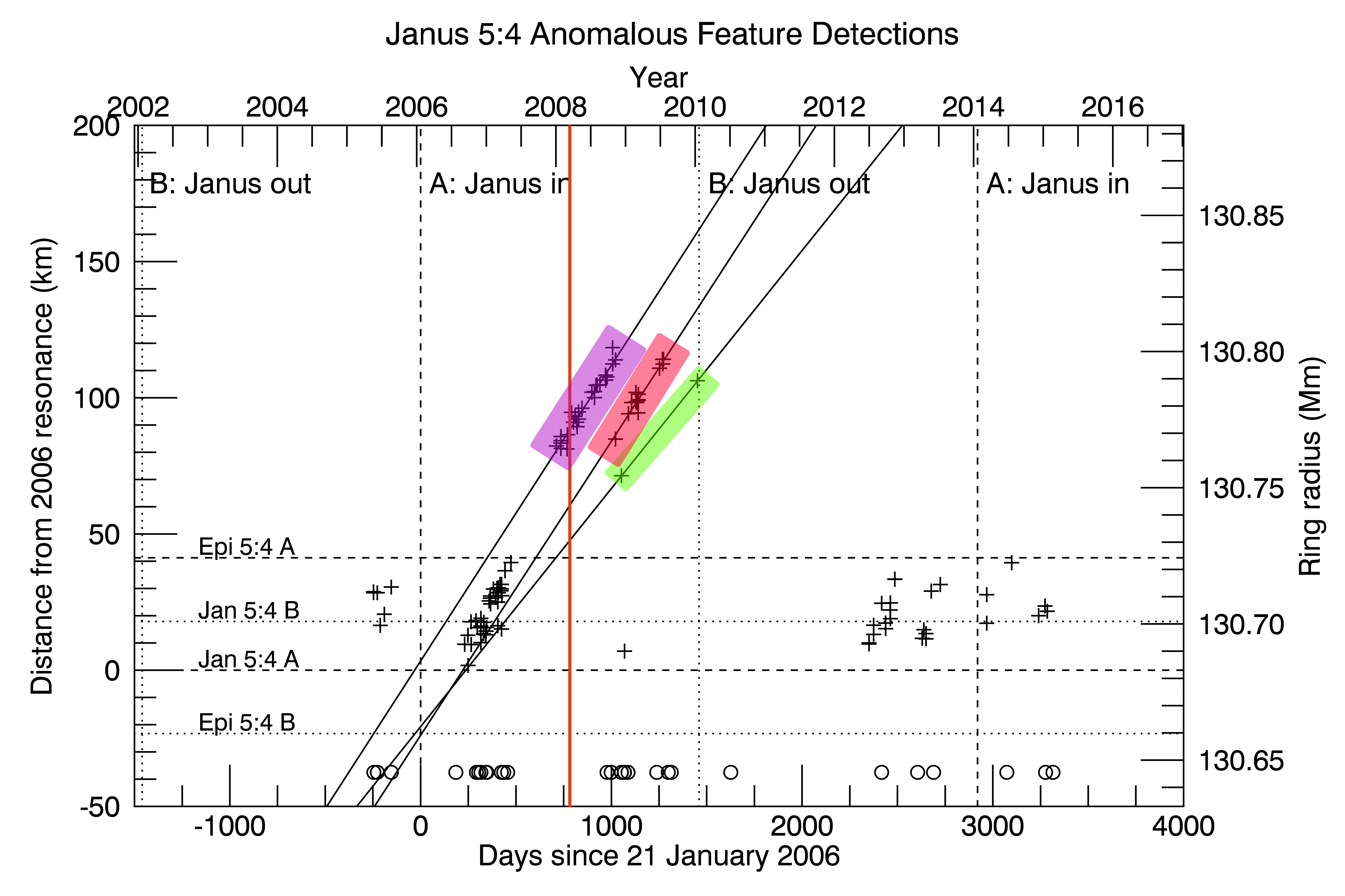}
\caption{The diagonal lines represent fits to the collections of detections indicated by the colored rectangles. The solid, vertical red line indicates the earliest possible date for the Janus-in Janus wave to reach the Janus-in Epimetheus wave. Colors do not correspond to those of figure~\ref{fig::j43-annotated}. The slopes of the lines are tabulated in table~\ref{tab::line_fits}.}
\label{fig::j54-lines}
\end{center}
\end{figure}

\begin{table}
\begin{center}
\begin{tabular} { c c c }
Resonance & Region & Velocity (km/yr)\\
\hline
Janus 6:5 & Purple (left) & 37.6$\pm$3.9\\
& Red (middle) & 44.4$\pm$2.6\\
& Green (right) & 39.8$\pm$1.6\\
Janus 5:4 & Purple (left) & 39.6$\pm$0.85\\
& Red (middle) & 39.4$\pm$1.5\\
& Green (right) & 31.9$\pm$1.5\\
\end{tabular}
\caption{Feature velocities implied by the lines fit in figures~\ref{fig::j65-lines}~and~\ref{fig::j54-lines}. Regions correspond to those figures. Uncertainty is 1 $\sigma$ with equal weighting of points. Note that the Janus 5:4 green region contains only two points.}
\label{tab::line_fits}
\end{center}
\end{table}

The first notion is that after the Janus-in swap, it takes some time for the Janus-in Janus wave to propagate outwards to the Janus-in Epimetheus wave, at which point interference as described above for region~1 occurs. The timescale for this propagation is 2.1 years for the Janus 5:4 wave and 2.0 years for the Janus 6:5 wave. Figures~\ref{fig::j65-lines}~and~\ref{fig::j54-lines} show fits to each possible cluster of anomalous detections. This is necessary because each launched solitary wave may not have the same velocity. This hypothesis is plausible for the right-most (green) group of detections, however we must recall that the new density waves will not begin forming on exactly 21 January and thus these timescales represent a lower limit.

A second hypothesis is that a single solitary wave splits into two or more waves subsequent to being launched, a behavior of solitons that has been observed in other contexts \citep{Hammack1974}. This is most plausibly demonstrated by the fact that the left-most two (purple and red) fitted lines clearly bracket the initial sequence of detections in figures~\ref{fig::j65-lines}~and~\ref{fig::j54-lines}. Because these fitted lines are parallel within their uncertainties (indicating equal velocities, see table~\ref{tab::line_fits}), this must have been a one-time event, not a slow diverging of two initially-launched waves.

Neither of these explanations are fully satisfying, most notably because they do not account for why a single occultation rarely observes more than one anomalous feature. Each occultation track represents a vertical line on these figures; thus, there are many instances in which, for example, a blue-feature detection is noted in an occultation but not a red or green feature.

\begin{figure}
\begin{center}
\includegraphics[width=1\textwidth]{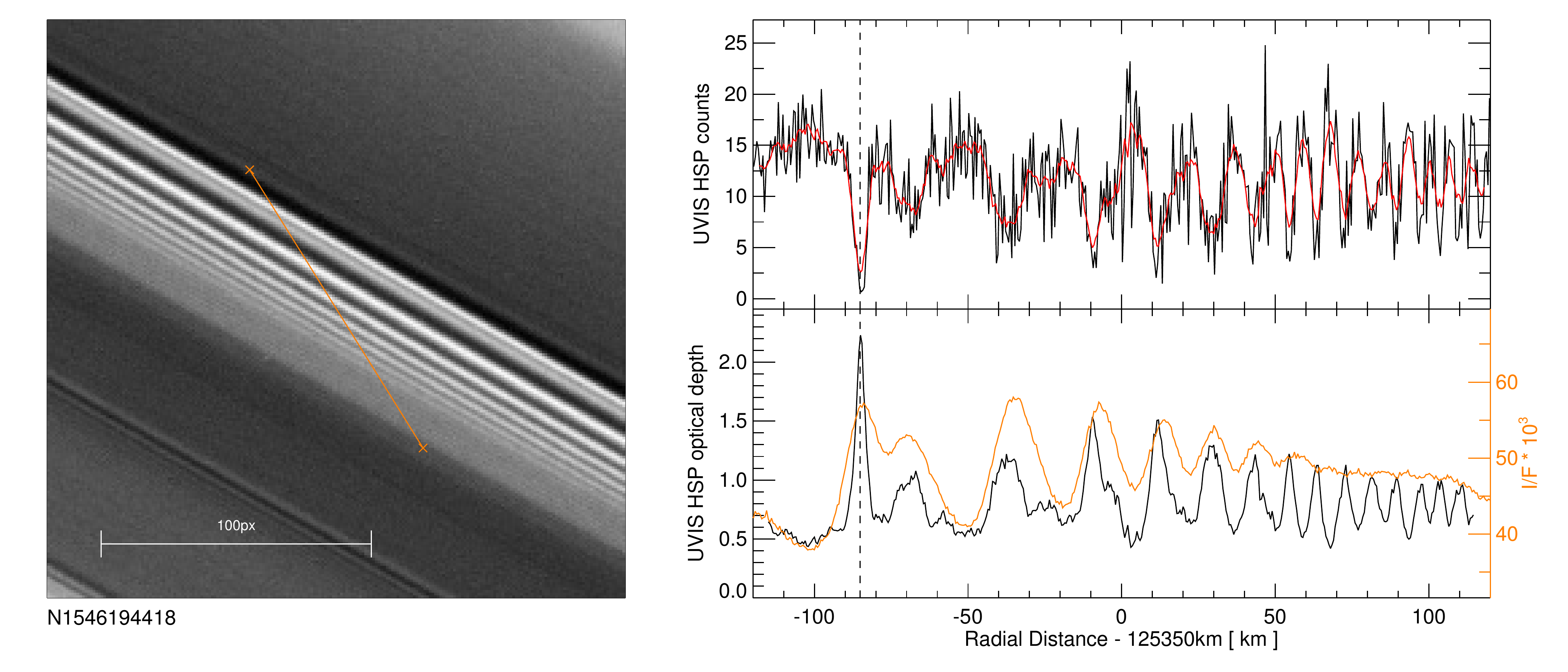}
\caption{An ISS narrow-angle camera observation of the Janus 4:3 spiral density wave taken simultaneously with a UVIS HSP occultation. \textit{Left}: The UVIS occultation track overlaid on the image. This image has a resolution of approximately 4 km/px and was taken approximately one minute prior to the occultation track passing through the anomalous feature. \textit{Top right}: The HSP occultation sampled to 500 m (black line) and 4 km (red line). The dashed line passes through the peak of the anomalous feature. \textit{Bottom right}: The I/F measured in a 3-px-wide profile along the UVIS track in the NAC image (orange line) and the HSP counts sampled to 4 km (black line).}
\label{fig::targeted-1}
\end{center}
\end{figure}

\subsection{Janus 2:1 region}
Unlike the other resonances examined in this work, the Janus 2:1 lies in the B ring. Physical conditions in this region may be different from those in areas of the A ring and we correspondingly observe a different manifestation of the feature. Nevertheless, we believe the generative mechanism in the Janus 2:1 region is broadly the same as described above. Instead of traveling at approximately twice the group velocity of the local spiral density wave, the feature in this region propagates at a commensurate speed. Because this velocity is much slower (see table~\ref{tab::velocities}), the solitary wave should not propagate out of the examined region within four years. Indeed, we see ``copies" of the wave that are visible more than 100 km from the resonance in figure~\ref{fig::quad_plot}. That multiple copies persist is consistent with \citet{Horn1996}, who observed a similar phenomenon for wakes generated by the moon Pan. However, the velocity necessary to achieve the observed separation in four years is not consistent with implied velocities of either the upper or lower sequences observed during the 2006-2010 period. The velocity required, 32 km/yr, is substantially larger than that implied by fits to any sequence of detections in the Janus 2:1 region. The fact that the computed feature velocity is bracketed by the density wave group velocities estimated for two different surface mass densities might lend support to the variable surface mass density hypothesis of \citet{Hedman2016}.

\subsection{Observations with the Imaging Science Subsystem}
Continuous features like the ones described above should be easier to detect in ISS imagery than a large population of discrete objects. We undertook a successful search of archival imaging data for possible detections of either the solitary wave or the interference patterns in the spiral density wave in images captured simultaneously with occultations. Figure~\ref{fig::targeted-1} illustrates a feature detection within a high-resolution targeted \textit{Cassini} narrow-angle camera (NAC) image of the Janus 4:3 region, one of four such images found. A similar image was found for each of the Janus 2:1, Janus 5:4, and Janus 6:5 regions.

The radial locations of these features fall along the predicted path of the solitary wave, providing a confirmation of this phenomenon independent of HSP stellar occultations. Moreover, each image shows that the feature extends at least as wide as the NAC field of view, a distance of 5446 km for the image in figure~\ref{fig::targeted-1}. This is compelling evidence in favor of a single continuous wave rather than a collection of discrete objects.

\section{Conclusion}
It is clear that the orbital swap of Janus and Epimetheus has a substantial effect on the rings. \citet{Tiscareno2006b} have demonstrated that it accounts for the unusual morphology of second-order, linear spiral density waves and \citet{ElMoutamid2016} have shown it substantially alters the shape of the outer edge of the A ring. Our investigation reveals that in Saturn's strongest density waves, the effect is even more dramatic. Nonlinear interference between the waves generated at the pre- and post-swap resonance locations results in the formation of a solitary wave. In the A ring, this wave propagates outward at about twice the group velocity of the local spiral density waves and in the B ring the propagation velocity is commensurate to the density wave group velocity.

Occultations observed by \textit{Cassini} throughout 2016 and images captured until the end of the mission should reveal the next iteration of this cycle and provide a longer baseline over which to understand the phenomenon. A future model describing the behavior of the solitary wave may also allow for an independent estimation of the ring surface mass density.

Finally, these observations have potential applications in understanding the physics of protoplanetary and accretion disks, where the migration of massive bodies has been theorized. Although we may generally lack the capability of detecting these moving bodies, the effects they render onto the neighboring disk of particles and gas may reveal their motion indirectly.

\section{Acknowledgements}
The authors wish to thank Linda Spilker and an anonymous reviewer for helpful comments on this manuscript.

\bibliographystyle{plainnat}
\bibliography{rings.bbl}

\end{document}